\documentclass[aps,prb,twocolumn,showpacs]{revtex4}

\usepackage{graphicx}
\usepackage{amssymb}

\begin{document}

\title{Point defect dynamics in bcc metals}

\author{J\"{o}rg Rottler, David J.~Srolovitz, and Roberto Car}
\affiliation{Princeton Institute for the Science and Technology of
Materials (PRISM), Princeton University, Princeton, NJ 08544}

\date{\today}

\begin{abstract}
We present an analysis of the time evolution of self-interstitial atom
and vacancy (point defect) populations in pure bcc metals under
constant irradiation flux conditions. Mean-field rate equations are
developed in parallel to a kinetic Monte Carlo (kMC) model.  When only
considering the elementary processes of defect production, defect
migration, recombination and absorption at sinks, the kMC model and
rate equations are shown to be equivalent and the time evolution of
the point defect populations is analyzed using simple scaling
arguments. We show that the typically large mismatch of the rates of
interstitial and vacancy migration in bcc metals can lead to a vacancy
population that grows as the square root of time. The vacancy cluster
size distribution under both irreversible and reversible attachment
can be described by a simple exponential function. We also consider
the effect of highly mobile interstitial clusters and apply the model
with parameters appropriate for vanadium and $\alpha-$iron.
\end{abstract}

 
\maketitle 

\section{Introduction}
Many properties of metals depend crucially on the type and
concentration of defects that perturb the ideal crystal structure. Of
these, the simplest are point defects, such as self-interstitials and
vacancies.  Since their formation energies are of order electron
volts, their equilibrium concentrations tend to be very low. They do
form in abundance, however, in radiation environments due to the
collisions between the irradiating species (electrons, heavy ions or
neutrons) and the atoms of the host crystal \cite{johnson}. When the
energy of the impinging particle is close to, but above, the
displacement threshold, the collision typically produces a single
Frenkel pair. Particles with higher kinetic energy, for instance
neutrons produced in fusion reactions, create collision cascades which
can produce not only Frenkel pairs, but an ensemble of mobile and
immobile self-interstitial and vacancy clusters of different
sizes. The ensuing evolution of the point defect distributions due to
diffusion \cite{flynn} determines the long time scale degradation of
the mechanical properties of the material, in addition to volume
swelling at large doses.

For body centered cubic (bcc) metals, such as $\alpha$-Fe, V and Mo,
molecular dynamics simulations have revealed a detailed microscopic
picture of the diffusive motion of individual defects and
vacancies\cite{delarubia97}. While the migration barrier $\Delta E_v$
for vacancy migration is rather high $(\sim0.5$ eV), both
self-interstitial atoms and small self-interstitial cluster are highly
mobile and easily diffuse along particular crystallographic directions
($\langle 111\rangle$-directions). In V, the lowest energy
self-interstitial configuration is a $\langle 111\rangle$-oriented
dumbell, which migrates easily with a crowdion transition state
\cite{han1}. However, this easy migration, with barriers as low as
$\Delta E_i\sim 0.02$ eV \cite{zepeda04}, leads to long
one-dimensional self-interstitial diffusional trajectories. The
self-interstitial dumbbell can change direction by rotating into other
$\langle111\rangle$-directions. The barrier associated with such
rotations $\Delta E_r$ is of the same order as $\Delta E_v$ (i.e.,
$\Delta E_r\gg\Delta E_i$).

In $\alpha$-Fe, by contrast, the ground state of the dumbbell
self-interstitial is the $\langle 110\rangle$ configuration, and
accessing the easy-glide $\langle 111\rangle$ configuration requires
overcoming a qualitatively similar rotational barrier that separates
the $\langle 110\rangle$ from the $\langle 111\rangle$ configurations
\cite{wirth97}.  Although the migration barriers for the crowdion
mechanism in $\alpha$-Fe are also believed to be very small $(<0.04$
eV), the observed effective migration barrier (including rotation into
the $\langle 111\rangle$ orientation and migration along the $\langle
111\rangle$-direction) are higher than in V
\cite{marian01,soneda01}. Similar arguments apply to Mo and Ta. Both
diffusion mechanisms imply that interstitial transport in bcc metals
takes place in the form of long one-dimensional trajectories with
occasional directional changes that become more frequent as the
temperature increases. In fcc metals such as Cu, for instance,
simulations showed that self-interstitial diffusion occurs through
much more conventional, isotropic diffusion mechanisms
\cite{osetsky01}. In most metals, however, large separations in time
scale between self-interstitial and vacancy motion exist.

The intent of the present study is to illuminate the consequences of
the intriguing microscopic diffusion mechanisms in bcc metals on the
evolution of the point defect population using simple, but readily
generalizable models. Our models shall minimally include local
self-interstitial/vacancy mutual annihilation and absorption into
(unsaturable) sinks which, in a real system, are in the forms of
dislocations or grain boundaries. Together with the continuous
production of point defects, these processes form the fundamental
events in metals under irradiation conditions.

On the continuum level, point defect dynamics can be described by a
set of kinetic master or rate equations that treat the point defects
as a continuous density whose temporal evolution is governed by
various gain and loss processes. These equations do not take into
account spatial correlations and are only analytically tractable in
the simplest situations. Although we focus the discussion specifically
on bcc metals, the continuum theory makes no reference to the
underlying crystal structure and could be readily applied to fcc
metals as well. In many cases, single interstitials and
vacancies combine with defects of the same type to form stable
clusters. The evolution of the cluster size distribution and other
microscopic variables is more conveniently studied in a particle based
model that explicitly represents defects and their diffusion
mechanisms (translation, rotation) on a lattice.  The competition
of events occuring with different rates can then be followed using a
kinetic Monte Carlo (kMC) scheme.

Both master equation and Monte Carlo approaches have frequently been
employed to illuminate the physics of damage evolution in irradiated
materials. Starting from the elementary processes mentioned above
\cite{brailsford72}, steady-state rate equations have been used to
study the effects of preferential absorption of self-interstitial
atoms at sinks (absorption bias) \cite{woo96,singh97}, the formation
of interstitial cluster during the cascade phase (production bias)
\cite{woo96,singh97,golubov00,trinkaus00}, and the one-dimensional
motion of these clusters in bcc metals
\cite{singh97,golubov00,trinkaus00}. kMC models were first used to
evolve the primary damage state obtained from ns-long MD simulations
to macroscopic time scales, but have increasingly been used to study
the evolution of defect structure during continuous irradiation in
copper \cite{heinisch96,heinisch97,heinisch99}, vanadium \cite{alonso}
and iron \cite{soneda03}.

Although many treatments have included a high level of atomistic
detail from the start, we begin by analyzing the simplest situation
that includes only production of Frenkel pairs, point defect
recombination and absorption at sinks in Section
\ref{simple-sec}. Even though this simplified case has been studied
many times before \cite{brailsford72}, we shall see that the inclusion
of specific features of bcc metals leads to a surprisingly rich
evolutionary picture. This approach allows us to gradually increase
the level of complexity of the model in a verifiable, controlled
manner. We introduce first interactions between vacancies that lead to
vacancy cluster formation in Section \ref{vaccluster-sec} and then
interactions between interstitials in Section \ref{intclust-sec}. In
particular, we shall discuss how interstitial cluster mobility affects
vacancy cluster formation. While the discussion in the text is
applicable to bcc metals broadly, we apply our findings to the
specific cases of V and $\alpha$-Fe in Section \ref{real-sec}.

\section{Point defect dynamics in simple situations}
\label{simple-sec}
\subsection{Atomistic details and kMC model}
\label{model-sec}

The physics of radiation damage evolution is governed by the
production of defects due to irradiation and their subsequent
elimination from the population through diffusional processes. Typical
values for the defect production rate $\sigma F$ range between
$10^{-3}$ dpa/s, in ion irradiation experiments \cite{iwai96}, and
$10^{-10}$ dpa/s, for neutron irradiation \cite{heinisch99}, where $F$
is the irradiation flux and $\sigma$ is the cross-section. A
``displacement per atom'' (dpa) refers to the production of one
Frenkel defect pair per lattice site.

Self-interstitial transport is composed of two parts, a
one-dimensional diffusive motion along one of the 4 distinct $\langle
111 \rangle$ directions and dumbbell rotations from one $\langle 111
\rangle$ direction to another. The temperature dependence of the
(one-dimensional) diffusivity $D_i$ is usually described through the
Arrhenius form in most cases except for vanadium, where the unusually
low activation barrier can lead to more complicated behavior at higher
$T$ (see Ref.~\onlinecite{zepeda04}).  For $T<600K$, however, the
interstitial diffusivity is well described by the Arrhenius expression
$D_i/a^2=f\nu_i\exp{[-\Delta E_i/k_BT]}$, where $\Delta E_i$ is the
activation energy barrier, $\nu_i$ an attempt frequency and
correlations are expressed through the correlation factor $f$ ($f=1$
when the diffusion is uncorrelated). The temperature dependence of the
rotation rate $\gamma_r$, by contrast, is always of the Arrhenius
form, $\gamma_r=\nu_r\exp{[-\Delta E_r/k_BT]}$, where $\Delta E_r$ the
characteristic rotation barrier and $\nu_r$ is an attempt
frequency. Values for both $\nu_i$ and $\nu_r$ range between
$10^{12}s^{-1}$ and $10^{13}s^{-1}$. Finally, vacancies diffuse
three-dimensionally with rate $D_v/a^2=\nu_v\exp{[-\Delta E_v/k_BT]}$,
where $\Delta E_v$ and $\nu_r$ are the activation barriers and attempt
frequencies, respectively. The defects can undergo two basic
reactions: recombination once an interstitial and vacancy defect are
within a certain ``recombination volume'', or absorption at sinks.
Typical dislocation densities in metals are of the order $10^{12} -
10^{14}\,{\rm m}^{-2}$, which translates into dislocation sink
densities (per lattice site) of $n_s \sim 10 ^{-4}-10 ^{-6}$.

Since the relevant time scales (ns-sec) for the evolution of the point
defect distributions far exceed those of molecular dynamics
simulations, we employ a coarse grained description, in which we
represent vacancies, dumbbell interstitial configurations and sinks as
pointlike objects that occupy ideal lattice sites of a bcc lattice
with lattice parameter $a$.  In order to mimic constant irradiation
conditions, new pairs of defects (self-interstitial and vacancy) are
introduced randomly onto defect-free lattice sites with rate $\sigma
F$. The microscopic diffusion process is replaced by instantaneous
hops of point defects to vacant neighboring lattice sites with rates
$D_i/a^2$ and $D_v/a^2$, respectively.  A self-interstitial is
constrained to forward-backward hops along one of the four $\langle
111 \rangle$ directions, but can also rotate to another $\langle 111
\rangle$ direction with rate $\gamma_r$. Vacancies diffuse
isotropically.  A self-interstitial or vacancy recombines if it finds
itself next to a vacancy or self-interstitial, respectively, or is
absorbed if one of its 8 neighboring sites contains a sink.  At each
time step, an event is chosen according to its probability and then
executed. Time is advanced according to the usual continuous time
algorithm \cite{binder92}, where the time increment is chosen from an
exponential distribution.

\subsection{Basic rate equations}
The elementary processes described in the kMC model can,
alternatively, be described within a rate equation formalism. Before
the advent of large scale computer simulations, this method
represented the only viable theoretical approach for simulating long
time radiation damage evolution.  The rate equations can be solved by
direct numerical integration or direct analysis in limiting cases.  In
the minimal model discussed above, the time evolution of the number
densities of the interstitals $n_i(t)$ and vacancies $n_v(t)$ is given
by the coupled nonlinear equations \cite{sizmann68,brailsford72,sizmann74}
\begin{eqnarray}\nonumber
\frac{dn_i}{dt}&=&\sigma
F-\kappa_v\omega_{iv} n_i-\kappa_i\omega_{vi} n_v-\kappa_{is}\omega_{is} n_i\\
\frac{dn_v}{dt}&=&\sigma
F-\kappa_v\omega_{iv}n_i-\kappa_i\omega_{vi} n_v-\kappa_{vs}\omega_{vs} n_v
\label{rate-eq}
\end{eqnarray}
Defect pairs are added to the population at a rate proportional to the
particle flux $F$ and a cross-section $\sigma$. Loss can occur through
a diffusing interstitial recombining with a vacancy with rate
$\omega_{iv}$ and a diffusing vacancy recombining with an interstitial
with rate $\omega_{vi}$. The recombination rates are all proportional
to the diffusion constant of the moving defect, but depend on the
dimensionality of the diffusion process. $\kappa_i$ and $\kappa_v$ are
dimensionless capture numbers that represent the spatial extent of the
defects and their effective (possibly long ranged) interactions.
Losses can also occur through absorption at sinks with rates
$\omega_{is}$ and $\omega_{vs}$ and corresponding capture numbers
$\kappa_{is}$ and $\kappa_{iv}$ for interstitials and vacancies,
respectively.  An alternative representation of Eqs.~(\ref{rate-eq})
is to define sink strengths $k_x^2$ via the relation
$k_{x}^2D_{x}=\kappa_{x}\omega_x$, where the subscript $x$ refers to any
of the combination of indices used above.

The encounter rates of the defects are given by the number of distinct
sites visited by a random walker per unit time. Since the mean squared
displacement $\langle R^2 \rangle=l^2N$ for a random walk with step
length $l$, the number of sites visited is $s=[\langle R^2
\rangle/l^2]^{1/2}\sim N^{1/2}$ in one dimension. By contrast, a
detailed analysis of random walks on three-dimensional cubic lattices
shows that a random walker visits $\mathcal{O}(N)$ distinct sites
after $N$ hops.  For a given density of target sites $n$, the typical
collision time $\tau_c$ is given by the condition $D\tau_c/a^2=1/n$
(3D) and $(D\tau_c/a^2)^{1/2}=1/n$ (1D), from which we deduce the
encounter rates
\begin{equation}
\omega_{3D}\sim nD/a^2 \qquad {\rm and} \qquad \omega_{1D}\sim n^2D/a^2.
\end{equation}
In a similar manner, the capture numbers $\kappa$ are also affected by
the dimensionality of the random walk. Assuming ideal spherical
defects of linear dimension $r$, $\kappa \sim r/a$ for a 3D random
walk in a mean-field approximation \cite{brailsford81}, but in
general, capture numbers can also depend on spatial fluctuations and
dose. For a 1D random walk, the scaling with defect size becomes much
stronger \cite{barashev01}, $\kappa \sim (r/a)^4$.

These expressions apply to strictly 1D or 3D random walks. As
discussed in the Introduction, we encounter an intermediate case in
bcc metals, where rotations interrupt 1D random walks and lead to 3D
trajectories. The encounter rate of this random walk must, therefore,
be larger than the purely 1D case. For a rotation rate $\gamma_r$, the
random walker performs on average $D/\gamma_r a^2$ hops along a
particular direction before rotating into a new direction.  There are
$\gamma_r\tau$ of these segments during time $\tau$. The mixed 1D/3D
collision time thus follows from the condition $(D/\gamma_r
a^2)^{1/2}\gamma_r \tau_c=1/n$, which implies
\begin{equation}
\omega_{1D/3D}=\omega_{3D}\sqrt{\gamma_ra^2/D}
\label{om-mixed-eq}
\end{equation}
The mixed encounter rate $\omega_{1D/3D}$ scales like the 3D encounter
rate $\omega_{3D}$, but is reduced by the square root of the ratio of
the number of rotations to hops. This reaction rate has also been
derived in ref.~\onlinecite{barashev01}. This work views the kinetics
in the intermediate 1D/3D regime as an enhanced 1D reaction rate, but
the resulting expressions agree up to numerical prefactors. These
authors also showed that the size dependence of the mixed capture
number $\kappa \sim (r/a)^2$, and ref.~\onlinecite{trinkaus02}
provides an interpolation formula between the limiting cases using a
continuum description. While the encounter rates and reaction kinetics
of random walkers decrease when the dimensionality of the random walk
changes from three to one, the diffusivity does not, since the mean
squared displacement of a random walk of $N$ steps of length $l$ is
$\langle {\bf R}(N)^2\rangle=l^2N$ independent of dimensionality.

Note that these encounter rates are derived under the assumption of
collisions with a stationary target. The case of several colliding 1D
random walkers becomes equivalent to a 3D random walk because, from
the rest frame of a given walker, the other walkers appear to be
executing a 3D random walk. This case would be relevant for describing
the collisions of interstitials with each other, but not with the
vacancies or sinks.

Inserting these encounter rates into Eqs.~(\ref{rate-eq}), assuming
mixed 1D/3D encounter for diffusing interstitials, yields rate
eqations of the form
\begin{eqnarray}\nonumber
\frac{dn_i}{dt}&=&\sigma
F-n_i n_v(\kappa_v \sqrt{\beta}D_i/a^2-\kappa_i D_v/a^2)\\&-&\kappa_{is} n_s n_i \sqrt{\beta}D_i/a^2\\
\frac{dn_v}{dt}&=&\sigma \nonumber
F-n_i n_v(\kappa_v \sqrt{\beta}D_i/a^2-\kappa_i D_v/a^2)\\&-&\kappa_{vs} n_s n_v D_v/a^2,
\label{rate2-eq}
\end{eqnarray}
where $\beta=\gamma_ra^2/D_i$ is a dimensionless ratio that describes
the relative frequency of dumbbell rotations and diffusional hops.

\subsection{Simple scaling analysis}
We now specialize these results to the common case of bcc metals,
where $\gamma_ra^2/D_i\ll 1$ and $D_i/D_v\gg 1$. Before performing the
kMC simulations and solving the full rate equations numerically, it is
instructive to analyse the limiting behaviors of this system
\cite{sizmann68}. Inititally, there are no defects in the metal, and
only the first term in the rate equations is important. In this regime
(regime I), defect densities increase linearly with time,
\begin{equation}
n_i^{\rm I}=n_v^{\rm I}=\sigma F t.
\end{equation}

Once a sufficient density of defects has been produced such that the
encounter times between defects becomes smaller than the time between
creation events, loss through recombination ($2^{nd}$ and $3^{rd}$
terms) becomes important \cite{sizmann64}. As we shall see below,
typical parameter ranges lead to a situation in which defect
recombination becomes important before sink loss.  The balance between
creation and recombination makes a steady state possible:
$dn_i/dt=dn_v/dt=0$. Ignoring the sink terms, one readily finds for
this Regime II
\begin{equation}
\label{regII-eq}
n_i^{\rm II}=n_v^{\rm
II}=\left(\frac{\kappa_v\sqrt{\beta}D_i+\kappa_iD_v}{\sigma
Fa^2}\right)^{-1/2}\sim\Gamma^{-1/2},
\end{equation}
which only depends on the dimensionless ratio
$\Gamma=\frac{(\sqrt{\beta}D_i+D_v)}{\sigma Fa^2}$. The crossover from
Regime I to Regime II occurs at time $t_{\text{I/II}}=F\sigma
t_{\text{I/II}} \simeq \Gamma^{-1/2}$, i.e., $t_{\text{I/II}}\sim
n_i^{-1}(D_i/a^2)^{-1}$. Note that this scaling regime is bound from
above by the following condition: if $n_v$ becomes so large that the
time between interstitial-vacancy encounters is on average shorter
than the time between two subsequent rotations (i.e., if
$\Gamma<\beta^{-1}$), the scaling of the encounter time becomes that
of a 1D random walk. In this case, the scaling of the steady state
defect density with $\Gamma$ changes from $n_i^{\rm II}=n_v^{\rm
II}\sim\Gamma^{-1/2}$ to $n_i^{\rm II}=n_v^{\rm II}\sim
\Gamma^{-1/3}$.  This regime is unlikely to be of experimental
relevance.

Eventually, loss through sinks becomes important and the steady state
Regime II ends. Clearly, there exists a terminal steady state (see
below) in which all loss terms in Eqs.~(\ref{rate-eq}) balance the
creation of defects. If $D_i\gg D_v$ as in the bcc metal case,
however, the system will initially loose mostly interstitials and very
few vacancies. This breaks the symmetry between interstitials and
vacancies and $n_v\gg n_i$.  The second steady state is, therefore,
reached through a transient Regime III with distinct
scaling. Subtracting the two rate equations (and neglecting vacancy
loss at sinks),
\begin{equation}
\frac{d(n_v -n_i)}{dt}\simeq\frac{dn_v}{dt}=\kappa_{is}n_s\sqrt{\beta}\frac{D_i}{a^2}n_i,
\end{equation}
where $n_s$ denotes the sink density. This equation allows us to
eliminate $n_i$ in the rate equation for $n_v$, so that
\begin{equation}
\frac{dn_v}{dt}\simeq\sigma F -\frac{\kappa_vn_v}{\kappa_{is}n_s}\frac{dn_v}{dt}.
\label{regIII-eq}
\end{equation}
Integrating Eq.~(\ref{regIII-eq}), we
obtain (to leading order) power-law growth of $n_v$ with time,
\begin{equation}
n_v^{\rm III}\simeq (\kappa_sn_{is}\sigma F t/\kappa_v)^{1/2}.
\end{equation}
In the case of extremely large sink densities $n_s$, the encounter
rates $\omega_{is}$ between interstitials and sinks would not be 3D as
assumed above, but 1D-like. In this case, $n_s$ would have to be
replaced by $n_s^2$, but again, such high densities are unrealistic.
The crossover time $t_{\text{II/III}}$ between Regimes II and III can
be obtained from the condition $\Gamma\simeq \kappa_{is}n_s\sigma F
t_{\text{II/III}}/\kappa_v$, which implies $t_{\text{II/III}}\sim
(\kappa_v/\kappa_{is}n_s)^{1/2}/(D_i/a^2)$. The crossover occurs at
constant time, independent of the defect densities.

If there are only sinks for one species of defect and no sinks for the
other, Regime III will simply continue. In all other cases, a final
steady state will occur when all loss terms are important. Setting
again $\frac{dn_i}{dt}=\frac{dn_v}{dt}=0$ as in
Ref.~\onlinecite{sizmann74}, we obtain a condition for steady state,
\begin{equation}
\label{regIV-1}
\frac{n_v^{\rm IV}}{n_i^{\rm IV}}=\frac{\sqrt{\beta}D_i}{D_v}=\frac{\kappa_{is}}{\kappa_{vs}}\alpha\sqrt{\beta},
\end{equation}
where we have introduced a third dimensionless ratio $\alpha=D_i/D_v$.
The interstitial and vacancy populations will, in general, be
different.  The steady state values are the roots of the quadratic
equation
\begin{eqnarray}
\label{regIV-2}
(n_v^{\rm IV})^2 \frac{\kappa_{vs}\Gamma}{\kappa_{is}\alpha\sqrt{\beta}}-n_v^{\rm IV}\frac{\kappa_{vs}n_sD_v}{\sigma Fa^2}=1.
\end{eqnarray}
Since typically $n_s\ll 1$, the linear term can be neglected relative
to the quadratic term, and we obtain the scaling of the vacancy
density in Regime IV,
\begin{equation}
n_v^{\rm IV}\sim \left(\frac{\Gamma}{\alpha\sqrt{\beta}}\right)^{-1/2}.
\end{equation}
The crossover time $t_{\text{III/IV}}$ follows again from the
condition $\kappa_{is}n_s\sigma F t_{\text{III/IV}}/\kappa_v\simeq
(\Gamma/\alpha\sqrt{\beta})^{-1}$, i.e.,~$t_{\text{III/IV}}\sim
(\alpha\sqrt{\beta}\kappa_v/\kappa_{is}n_s)^{1/2}/(D_i/a^2)$ is again
independent of $n_i$ or $n_v$.

The volume swelling is proportional to the excess vacancy population
in this regime, $S=n_v-n_i$. Note that the imbalance arises here due
to the very different diffusivities of the two types of defects. Upon
termination of the irradiation, both the $n_i$- and $n_v$-populations
would relax exponentially to zero in this simple model.

\subsection{Numerical integration and kMC}
\begin{figure}[t]
\includegraphics[width=8cm]{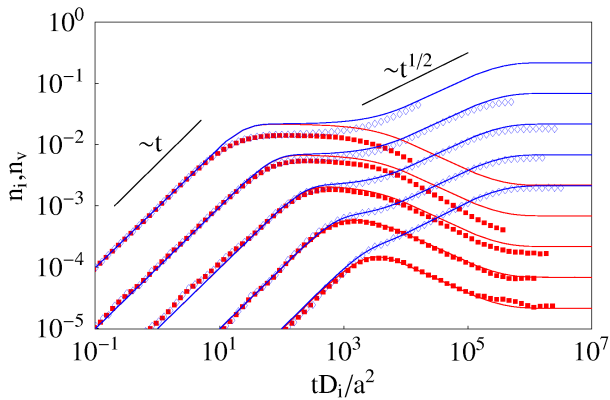}
\caption{\label{time-fig}Self-interstitial ($\blacksquare$), $n_i$,
and vacancy ($\lozenge$), $n_v$, densities as a function of time for
$\Gamma=10^2, 10^3, 10^4, 10^5,$ and $10^6$ ($\Gamma$ increases from
the top to bottom of the figure). Time is measured in units of the
inverse interstitial hopping rate $a^2/D_i$. The solid lines show the
result of direct numerical integration of Eqs.~(\ref{rate-eq}) with
$\kappa_v=\kappa_i=\kappa_{is}=\kappa_{vs}=21$, and the symbols correspond to the
results of the kMC simulations in a periodic simulation box. The
straight solid lines have slope $1$ and $1/2$.  }
\end{figure}

A full solution of the rate equations (\ref{rate-eq}) is only possible
numerically. In the following, we show a series of such numerical and
kinetic Monte Carlo simulation results for a representative choice of
parameters $D_i=1000 D_v$, $\gamma_r=0.01 D_i/a^2$, so that
$\alpha=1000$ and $\beta=0.01$ (see Section \ref{real-sec} for typical
experimental regimes). In addition, we set the density of sinks to
$n_s=10^{-4}$. Figure~\ref{time-fig} shows the evolution of the
average interstitial and vacancy density (symbols) based on the above
parameters obtained from the kMC simulation and numerical integration
of the corresponding rate equations (solid lines) using the expression
Eq.~(\ref{om-mixed-eq}) for the mixed encounter rates. The
dimensionless parameter $\Gamma$ was varied by changing the defect
creation rate $F\sigma$. As predicted by the scaling analysis, four
distinct regimes appear with increasing time. After first rising
linearly with time (Regime I), $n_i$ and $n_v$ reach the steady state
plateau of Regime II. Once loss through sinks becomes important
(Regime III), $n_v$ increases as $t^{1/2}$ while $n_i$
decreases. Finally, all curves reach the steady-state Regime IV, where
$n_i$ and $n_v$ are given by Eqs.~(\ref{regIV-1}) and
(\ref{regIV-2}). At the highest defect densities, the kMC simulations
where not carried out into the final steady state regimes, because the
computational effort becomes very large.

Excellent agreement between continuum theory and kMC model for the 3
largest values of $\Gamma$ is achieved by adjusting all capture
numbers to a single numerical constant. In the simple situation
examined here, kMC and rate theory are completely equivalent. Since we
have used $\omega_{\rm 1D/3D}$ in the rate equations, this agreement
also validates the scaling argument for the mixed 1D/3D encounter
rate.  For the two smallest $\Gamma$-values, we observe increasing
discrepancies between rate equations and kMC. As discussed above, we
expect a transition from the mixed 1D/3D encounter rates $\omega_{iv}$
to 1D dominated encounter kinetics when the interstitial density
becomes so high that they typically collide with vacancies or sinks
before rotating ($\Gamma<100$). This transition is not faithfully
reproduced by the rate theory which assumes only the limiting cases of
either 1D or mixed 1D/3D encounter rate scaling, but is properly
captured by the kMC which includes explicit 1D/3D trajectories.

Note also that with increasing $\Gamma$, Regime II begins to shrink
and Regime I crosses over directly into Regime III.  This happens
because as the defect densities decrease below the sink density, point
defect loss at sinks will become dominant before recombination plays a
significant role. Since the crossover time $t_{\text{I/III}}\simeq
\kappa_sn_s/\kappa_v\sigma F$ only depends on the point defect
production rate, a further decrease of the production rate will
eventually lead to a direct crossover from Regime I to Regime IV.  A
smaller sink density would push the appearance of Regimes II and III
to higher values of $\Gamma$.

\section{Vacancy cluster formation} 
\label{vaccluster-sec}
\subsection{Irreversible aggregation}
To this point, neglected interactions between vacancies. Experimental
vacancy-vacancy binding energies $E_b$ can be of order the single
vacancy migration energy.  This suggests that it is not unreasonable
to expect vacancies to form stable clusters or microvoids once they
meet. A relatively simple approximation for this situation is to
consider the limiting case of ``irreversible aggregation'', which
neglects any dissociation of a vacancy from a cluster and thus sets
$E_b=\infty$. The realistic situation, which includes finite binding
energies, can be viewed as an intermediate case between irreversible
aggregation and the case of $E_b=0$ of the previous section.

In the ``irreversible aggregation'' model, vacancies bind to form
stable divacancies, leaving the population of free vacancies, while
interstitials recombining with a divacancy recreate a single mobile
vacancy. Stable cluster grow or shrink under the influence of arriving
vacancies and interstitials. Denoting the number density of clusters
of size $m$ as $n_c(m)$, we can write the following set of rate
equations\cite{ghoniem89}:
\begin{eqnarray}\nonumber
\frac{dn_i}{dt}&=&\sigma
F-\kappa_i\omega_{vi} n_v\\\nonumber
&-&(\kappa_v\omega_{iv}+\kappa_{is}\omega_{is}+\sum_m\kappa_i^m\omega_{ic(m)})n_i\\\nonumber
\frac{dn_v}{dt}&=&\sigma
F-(\kappa_i\omega_{vi}+\kappa_{vs}\omega_{vs}+\sum_m\kappa_v^m\omega_{vc(m)}\\\nonumber
&-&2 \kappa_v \omega_{vv})n_v-(\kappa_v\omega_{iv}-\kappa_i^2\omega_{is(2)})n_i\\\nonumber
\frac{dn_c(m)}{dt}&=&(\kappa_v^{m-1}\omega_{vc(m-1)}-\kappa_v^{m}\omega_{vc(m)})n_v\\
&+&(\kappa_i^{m+1}\omega_{ic(m+1)}-\kappa_i^{m}\omega_{ic(m)})n_i ,
\label{rate-irr-eq}
\end{eqnarray}
where $n_v\equiv n_c(1)$ is now understood to refer to the free
vacancy density (or cluster of size 1). The additional terms in the
equations for the evolution of $n_i$ and $n_v$ account for
interstitial/vacancy-cluster encounters, vacancy/vacancy-cluster
encounters, vacancy-cluster nucleation and divacancy
decomposition. $\omega_{ic(m)}= n_c(m)\sqrt{\beta}D_i/a^2$,
$\omega_{vc(m)}= n_c(m)D_v/a^2$ and $\omega_{vv}=n_vD_v/a^2$ denote
the corresponding encounter rates. The last expression in
(\ref{rate-irr-eq}) describes the evolution of stable, immobile
vacancy clusters of size $m>1$ due to the diffusive arrival of
interstitials and vacancies. The hierarchy of rate equations
(\ref{rate-irr-eq}) is, in principle, amenable to a semi-analytical
treatment via numerical integration, if all constants $\kappa^m_{i/v}$ are
specified. However, such a solution requires termination of the set of
equations at a finite cluster size $m_{max}$, i.e., one imposes a
boundary condition $n_c(m_{max})=0$. The corresponding kMC model does
not require any ad hoc assumptions and takes all these processes
naturally into account.

Before examining this model in kMC, we can draw some premliminary
conclusions. In Regime I, only nucleation of divacancies will be
important. Since here the vacancy density grows linearly with time, we
predict the scaling regime
\begin{equation}
n_c(2)^I\sim(\sigma Ft)^3. 
\end{equation}
The divacancy density grows cubically with time. In the steady-state
Regime II, the vacancy cluster density is slaved to the free vacancy
density and, therefore, is constant as well. For $\alpha\gg1$, we
expect the total cluster density $n_c=\sum_m n_c(m)$ to be much
smaller than $n_v$ and $n_i$, so that the presence of vacancy
clustering does not yet strongly alter the interstitial and vacancy
population. Using Eq.~(\ref{regII-eq}), we find
\begin{equation}
n_c^{II}\simeq \Gamma^{-1/2}/(\alpha\sqrt{\beta}).
\end{equation}
Once past Regime II, the cluster density will begin to rise again as
$n_v$ increases. $n_v$ will then increasingly fall below the total
number of vacancies in the system $n_{vtot}$, because increasing
numbers of vacancies $\sum_{m>1} mn_c(m)$ become immobilized in
clusters. These clusters act as additional sinks for the diffusing
vacancies, but do not remove them from the system. The higher the
cluster density, the smaller the loss of vacancies through sinks and
interstitial recombination. We therefore expect that Regime III will
extend to larger times and the presence of the immobile vacancy
cluster delays the onset of the final steady-state Regime IV.

For the total number of vacancies $n_{vtot}$, we can write the rate
equations as
\begin{eqnarray}\label{nvtot-scaling-eq}
\frac{dn_{vtot}}{dt}&=&\sigma
F-\kappa_i\omega_{vi}n_v-\sum_{m=1}\kappa_i^{m}\omega_{ic(m)}n_i\\
\nonumber
\frac{dn_i}{dt}&=&\sigma
F-\kappa_i\omega_{vi} n_v-(\sum_{m=1}\kappa_i^m\omega_{ic(m)}-\kappa_{is}\omega_{is})n_i,
\end{eqnarray}
where we have neglected vacancy loss at sinks. This is the Regime III
situation discussed earlier. Subtraction of
Eqs.~(\ref{nvtot-scaling-eq}) and elimination of $n_i$, as in
Eq.~(\ref{regIII-eq}), leads to
\begin{equation}
\frac{dn_{vtot}}{dt}\simeq \sigma F -
\frac{\kappa_v}{\kappa_{is}n_s}\frac{dn_{vtot}}{dt}\sum_{m=1}\frac{\kappa_i^m}{\kappa_v}n_c(m).
\end{equation}
At relatively early times, when only small clusters are present,
$\sum_{m=1}\frac{\kappa_m}{\kappa_v}n_c(m)\simeq n_{vtot}$ and
$n_{vtot}$ will initially increase as $n_{vtot}\sim (\kappa_{is}n_s\sigma
F t)^{1/2}$. The subsequent behavior depends on the detailed form of
the capture numbers $\kappa_{is}$.

In the absence of vacancy clustering, Regime III ends when the vacancy
density has become so large that vacancy loss through sinks balances
interstitial loss through sinks. This condition, Eq.~(\ref{regIV-1}),
must hold for a final steady state to appear. In the present simple
model, this condition depends only on the ratio of the diffusivities,
but in a real system, Eq.~(\ref{regIV-1}) also depends on sink
concentrations and capture numbers, which may be different for
interstitial and vacancies. Eventually this condition can also occur
as a result of vacancy clustering, because vacancy clusters act as
additional sinks for the interstitials and remove them symetrically
from the system. At high cluster densities, the low mobile vacancy
concentration can therefore be compensated. In the steady-state regime
IV, we expect the hierarchy
$n_i^{IV}<n_v^{IV}<n_c^{IV}<n_{vtot}^{IV}$. Because of the later entry
into Regime IV, $S=n_{vtot}-n_i$ (the total amount of volume swelling)
will be larger than in the absence of clustering.
\begin{figure}[t]
\includegraphics[width=8cm]{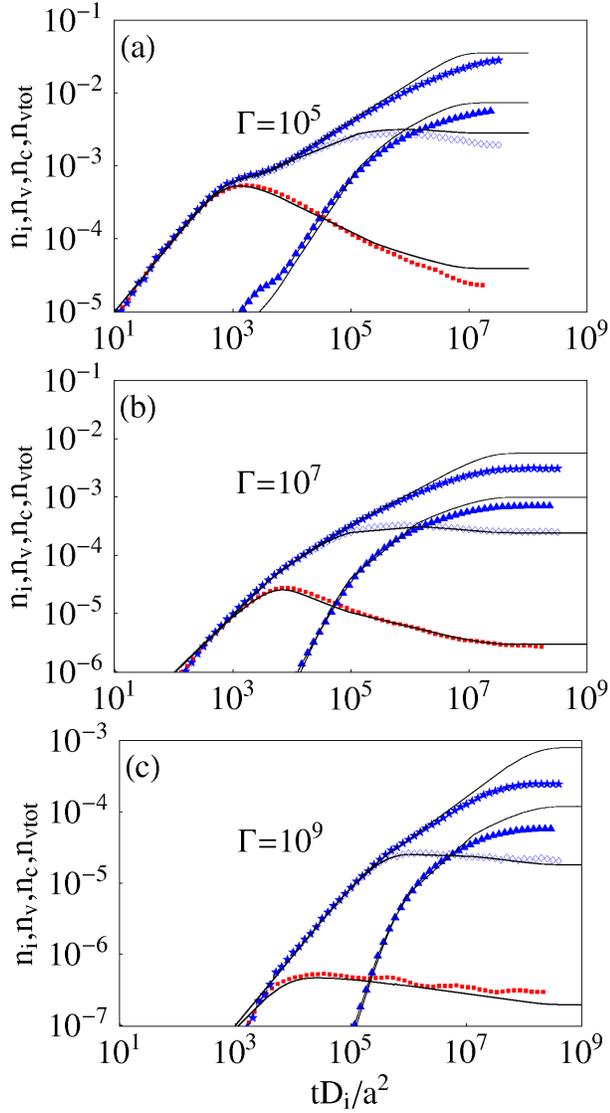}
\caption{\label{cluster-fig}Interstitial densities $n_i$ (
$\blacksquare$), free vacancy density $n_v$ ($\lozenge$) as well
as the total vacancy density $n_{vtot}$ ($\star$) and vacany cluster
density $n_{c}$ ($\blacktriangle$) as a function of time for (a)
$\Gamma=10^5$, (b) $\Gamma=10^7,$ and (c) $\Gamma=10^{9}$.  The thin
solid lines show the result of direct numerical integration of
Eqs.~(\ref{rate-irr-eq}) with $\kappa^m_i=\kappa_v^m=25$ and a $m_{max}=20$ (see
text).  }
\end{figure}

In Regime IV, we can obtain an approximate scaling relation for the
total cluster density. The density is determined by the competition
between nucleation of clusters out of the vacancy gas and
decomposition of divacancies due to arriving interstitials. Denoting
the fraction of clusters that represent divacancies as $f$, we can
write the rate equation
\begin{equation}
\frac{dnc^{{\text
IV}}}{dt}=\kappa_vn_v^2D_v/a^2-f(t)\kappa_i^2n_c^{\text
IV}n_i\sqrt{\beta}D_i/a^2,
\end{equation}
which implies $n_c^{\text
IV}=\kappa_vn_v^2/f\kappa^2_i\alpha\sqrt{\beta}n_i=\kappa_vn_v/f\kappa_2$,
where we have used Eq.~(\ref{regIV-1}). The cluster density is
therefore of order the free vacancy density with a numerical prefactor
that depends on the mean cluster size.

\subsubsection{Average cluster density}
The effect of irreversible vacancy aggregation on the defect dynamics
as seen through kMC is shown in Fig.~\ref{cluster-fig} for three
values of $\Gamma=10^5$, $\Gamma=10^7$ and $\Gamma=10^{9}$ with all
other parameters as before. In this figure, we also plot the total
cluster density $n_c$ and the total vacancy density $n_{vtot}$. In
Regimes I and II, $n_c$ and $n_i$ are nearly unchanged. The defect
density $n_c$ rises first $\sim t^3$ and then remains constant during
Regime II as discussed above. In Regime III, $n_v$ begins to fall
below the result from Fig.~\ref{time-fig} because of trapping of
vacancies into clusters. At the same time, $n_{vtot}$ rises with an
ideal $t^{1/2}$ law, and Regime III extends to larger times. $n_c$
also rises again due to new cluster nucleation events out of the
vacancy gas. All quantities reach constant values in regime IV. As one
might expect, the total number of vacancies in the system is higher
than in the absence of clustering. However, the general sequence of
regimes remains unchanged.

In order to gain additional insight into the growth dynamics, we also
numerically integrate the full set of rate Eqs.~(\ref{rate-irr-eq})
(solid lines in Fig.~\ref{cluster-fig}). This requires specification
of the capture numbers $\kappa^m$. In the simplest mean field model
for nucleation dynamics, the capture numbers do not depend on cluster
size $m$ and $\kappa^1_{i,v}=\kappa^m_{i,v}=const.$ (i.e., the point cluster
model). Interestingly, we find surprisingly good agreement between the
rate equations (solid lines in Fig.~\ref{cluster-fig}) and the full
kMC simulations when using this simple approximation.  This suggests
that for relatively small $m$, $\kappa^m$ is only very weakly
size-dependent. The rate equations faithfully reproduce the sequence
of regimes, but begin to show deviations at late times when larger
clusters become more prominent and the point cluster assumption is no
longer accurate. Here, $\kappa^m$ begins to show some size
dependence. The success of this comparison shows, however, that a
mapping between kMC and rate equations is possible even with void
nucleation and growth if the $\kappa^m$'s are accurately parametrized.

Note that in the present model, the volume swelling rate $dS/dt$ is
zero in Regime IV. This is possible because we have chosen the same
capture radii $\kappa_s$ for the sinks for both interstitials and
vacancies, i.e., the sinks are symmetric with respect to defect
type. $S\sim t$ would be valid, if, for example, the capture radius
for interstitials is larger than that for vacancies. This situation
can arise with the introduction of dislocation sinks (known
as``dislocation bias'') and has been invoked to explain unusually
large swelling rates (see also Section \ref{conc-sec}) at times later
than considered here \cite{woo96,singh97}.

\subsubsection{Cluster size distribution}
A full description of the evolution of the point defects in the
material includes a characterization of the cluster size
distribution. The growth of these clusters is the result of the
diffusive arrival of interstitial and vacancy defects at already
nucleated clusters. The full dynamics of this process is described by
the last equation in (\ref{rate-irr-eq}). Consider this equation in
the following simplified notation:
\begin{eqnarray}\nonumber
\frac{dn_c(m)}{dt}&=&n_v(\kappa_v^{m-1}n_c(m-1)-\kappa_v^m n_c(m))D_v/a^2\\
&+&n_i(\kappa_i^{m+1} n_c(m+1)-\kappa_i^{m} n_c(m))\sqrt{\beta}D_i/a^2.
\label{clustsize-eq1}
\end{eqnarray} 
If $n_i$ and $n_v$ change very slowly in time, the distribution
$n_c(m)$ is given by the steady state solution to
Eq.~(\ref{clustsize-eq1}), i.e., $dn_c(m)/dt=0$. This equation is
supplemented by the detailed balance condition
\begin{equation}
n_v\kappa_v^m n_c(m)D_v/a^2=n_i \kappa_i^{m+1}n_c(m+1)\sqrt{\beta}D_i/a^2,
\end{equation}
which should also hold for $m\ge 2$. From this condition, we can
deduce the distribution $n_c(m)$ by induction. Since $n_v\equiv
n_c(1)$, we have $n_c(2)=n_v^2\kappa_1/(\kappa^2_i\alpha\sqrt{\beta}
n_i)$ and consequently for all $m>1$
\begin{equation}
n_c(m)=\frac{n_v^m}{(\alpha\sqrt{\beta} n_i)^{m-1}}\frac{\prod_{l=1}^{m-1}\kappa_v^l}{\prod_{l=2}^m\kappa_i^l}.
\label{ncdist-eq}
\end{equation}

\begin{figure}[t]
\includegraphics[width=8cm]{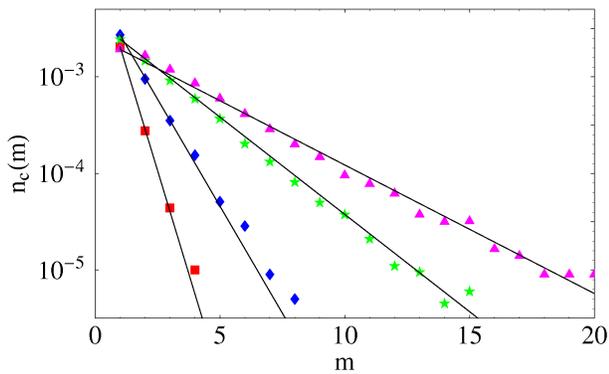}
\caption{\label{clusterdist-fig} Plot of the cluster size distribution
$n_{c}(m)$ found from the kMC simulations with $\Gamma=10^5$ of
Fig.~\ref{cluster-fig}(a) at four different times $tD_i/a^2=3\times
10^4$ $(\blacksquare)$, $3\times 10^5$ $(\blacklozenge)$, $3\times
10^6$ $(\star)$, and $3\times 10^7$ $(\blacktriangle)$ in Regimes III
and IV. The straight lines correspond to
$n_v(t)^m/(\alpha\sqrt{\beta}n_i(t))^{m-1}$.}
\end{figure}

Equation~(\ref{ncdist-eq}) is tested against the kMC results in
Fig.~\ref{clusterdist-fig}, where we show the cluster size
distributions at four different times. All curves fall along straight
lines in a semilogarithmic plot, and the slope decreases with
increasing mean cluster size. A comparison with Eq.~(\ref{ncdist-eq})
requires, again, knowledge of the capture numbers $\kappa^m$. As in
the previous section, use of the point cluster approximation
$\kappa^l_{i,v}/\kappa^1_{i,v}=1$ yields excellent agreement
between between the kMC data and Eq.~(\ref{ncdist-eq}) upon inserting
the values for $n_i(t)$ and $n_v(t)$ at the appropriate times.

Note that the distributions shown in Fig.~\ref{clusterdist-fig} are
not peaked, i.e., the frequency with which different clusters appear
decrease with increasing cluster size and single vacancies occur most
frequently. This situation is in sharp contrast to other cluster
growth situations as, e.g., found in submonolayer island growth during
vapor deposition \cite{amar95}, where the distribution $n_c(m)$ peaks
at the mean cluster size $\langle m \rangle$. In this problem,
however, clusters cannot shrink, since vacancies are absent. One
expects Eq.~(\ref{ncdist-eq}) to hold as long as the assumption of
quasi-stationary values for $n_i$ and $n_v$ is valid.
\begin{figure}[t]
\includegraphics[width=8cm]{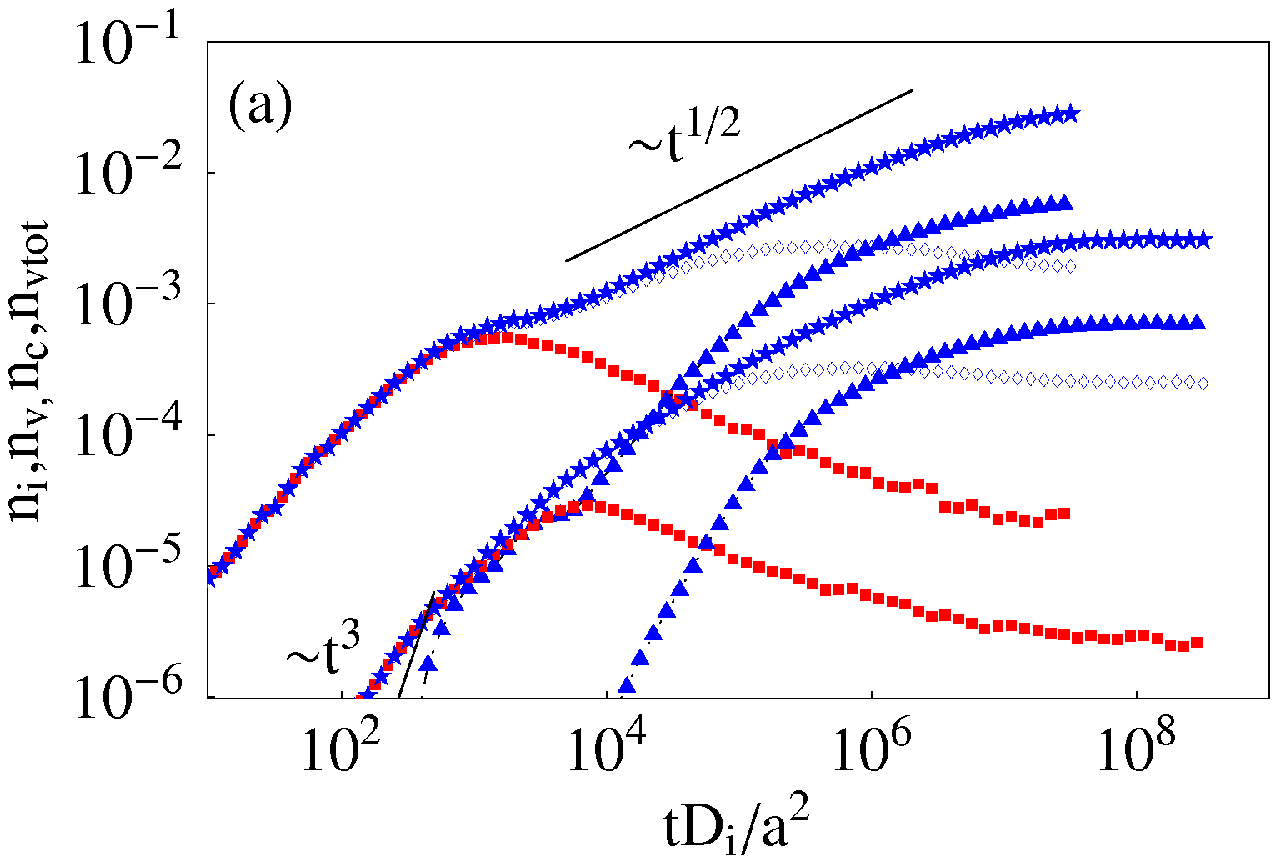}
\includegraphics[width=8cm]{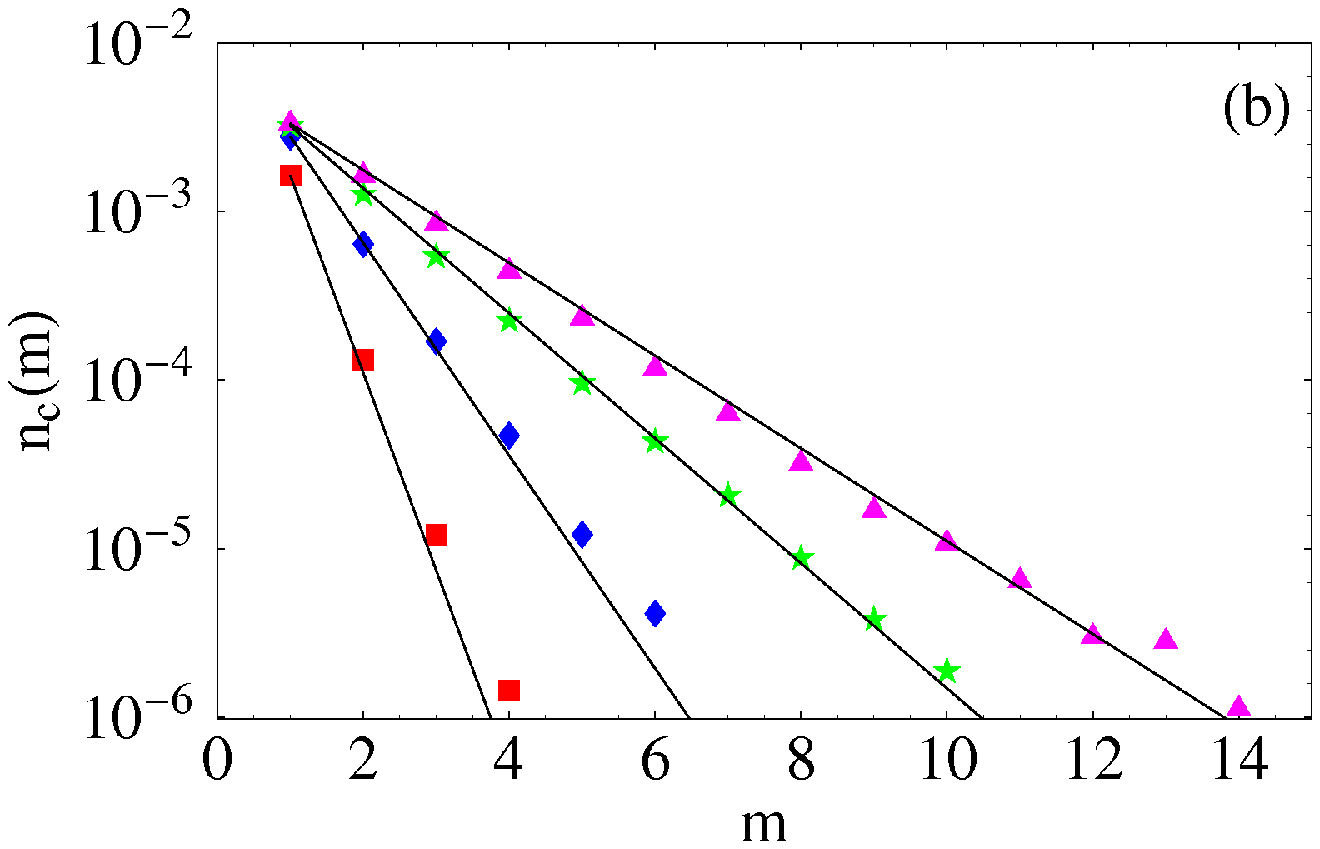}
\caption{\label{clusterrev-fig}(a) Interstitial density $n_i$
($\blacksquare$), free vacancy density $n_v$ ( $\lozenge$) as
well as the total vacancy density ($(\blacktriangle)$) and vacancy
cluster density $n_{c}$ $(\star)$ as a function of time for
$\Gamma=10^5$, and $10^{7}$ for reversible aggregation, where
$\delta=0.01$. (b) Cluster size distribution $n_{c}(m)$ for the case
of $\Gamma=10^5$ at four different times in Regimes III and IV. The
straight lines correspond to
$n_v(t)^m/(\alpha\sqrt{\beta}n_i(t)+\kappa_d^1/\kappa_i^1\alpha^\prime)^{m-1}$
with $\kappa_d^1/\kappa_i^1=0.25$.}
\end{figure}

\subsection{Reversible attachment}
The above discussion immediately raises the question of whether the
results for $n_c(t)$ and its size distribution Eq.~(\ref{ncdist-eq})
survive in the more realistic situation of reversible cluster growth,
i.e., vacancies have a finite probability to attach to and to leave
the cluster. Let us introduce ``detachment rates'' from a cluster of
size $m$, $\gamma_{det}(m)/a^2$ for all vacancies, independent of
their local environment. Equation~(\ref{clustsize-eq1}) needs to be
generalized by the addition of two terms,
\begin{eqnarray}\nonumber
\frac{dn_c(m)}{dt}&=&n_v(\kappa_v^{m-1}n_c(m-1)-\kappa_v^mn_c(m))D_v/a^2\\
&+&n_i(\kappa_i^{m+1}n_c(m+1)-\kappa_i^mn_c(m))\sqrt{\beta}D_i/a^2,\\
&+&(\kappa^{m+1}_d n_c(m+1)-\kappa^{m}_dn_c(m))\gamma_{det}(m)/a^2,
\label{clustsize-rev-eq}
\end{eqnarray}
where $\kappa^m_d$ represent ``detachment numbers'' in analogy to the
capture numbers $\kappa^m_{i,v}$. Consequently, the condition for detailed
balance now reads
\begin{eqnarray}\nonumber
\kappa_v^{m}n_vn_c(m)D_v/a^2&=&\kappa_i^{m+1}n_in_c(m+1)\sqrt{\beta}D_i/a^2\\
&+&\kappa^{m+1}_dn_c(m+1)\gamma_{det}(m)/a^2
\end{eqnarray}
and Eq.~(\ref{ncdist-eq}) generalizes to 
\begin{equation}
n_c(m)=\frac{\prod_{l=1}^{m-1}\kappa_v^l
n_v^m}{\prod_{l=2}^{m}(\kappa_i^l\alpha\sqrt{\beta}
n_i+\kappa_l^d\alpha^\prime(m))},
\label{ncdist-rev-geneq}
\end{equation}
where $\alpha^{\prime}(m)=\gamma_{det}(m)a^2/D_v$. We see that the
general form of the distribution is the same, but the prefactor
changes due to the additional growth and shrinkage probabilities.
From a statistical point of view, interstitial arrival and vacancy
detachment are equivalent. Eq.~(\ref{ncdist-rev-geneq}) can be easily
evaluated for any functional form of the size-depedent capture numbers
and detachment rates.  In the point cluster approximation, where all
$\kappa^m_{i,v,d}/\kappa^1_{i,v,d}=1$ and
$\alpha^\prime=\alpha^{\prime}(m)$ is size-independent, it predicts an
exponential distribution as in the case of irreversible attachment,
\begin{equation}
n_c(m)=\frac{n_v^m}{(\alpha\sqrt{\beta} n_i+\alpha^\prime\kappa^1_d/\kappa_i^1)^{m-1}}.
\label{ncdist-rev-eq}
\end{equation}

Figure~\ref{clusterrev-fig} presents a numerical investigation of
reversible vacancy cluster growth using kMC. Although detachment rates
may be cluster size dependent in general, we only introduce one
size-independent rate $\gamma_{det}/a^2=0.01D_{v}/a^2$ for vacancy
detachment for simplicity. This model would be most relevant for
faceted cluster shapes with one dominant detachment rate from the
faces. The other parameters are that of Fig.~\ref{cluster-fig}. We see
in Fig.~\ref{clusterrev-fig}(a) that the cluster density behaves in a
qualitatively similar manner as in the irreversible case, but $n_c$ is
reduced and the final steady state is reached at earlier times. As in
the irreversible case, the cluster size distribution $n_c(m)$ shown in
Fig.~\ref{clusterrev-fig}(b) is well described by
Eq.~(\ref{ncdist-rev-eq}). The present discussion is of course only
relevant when void coarsening can be neglected.
 
\begin{figure}[t]
\includegraphics[width=8cm]{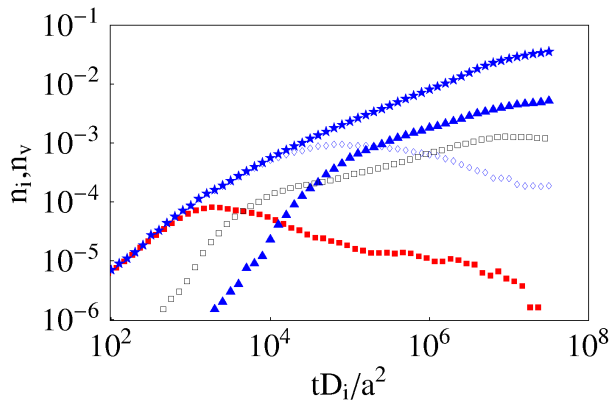}
\caption{\label{intclustimm-fig} Interstitial densities $n_i$ (
$\blacksquare$), free vacancy density $n_v$ ($\lozenge$), total
vacancy density $n_{vtot}$ ($\blacktriangle$), vacany cluster density
$n_{c}$ ($\star$) and density of immobile interstitial cluster
$(\square)$ as a function of time for $\Gamma=10^6$.  }
\end{figure}

\section{Interstitial interactions}
\label{intclust-sec}
One of the fascinating aspects of point defect dynamics in bcc metals
is the high mobility of interstitial clusters \cite{osetsky01}. The
interaction between single interstitial atoms is even stronger than
that between vacancies, and the interstitial clusters, such as
dislocation loops \cite{soneda98}, are stable at all relevant
temperatures.  Unlike the immobile vacancy cluster, however, the
interstitial cluster migrate easily for clusters with up to 50 or 100
interstitial atoms \cite{soneda01,soneda03}. We now modify the
preceding analysis to account for this behavior.

In this analysis, we return to the irreversible aggregation limit,
i.e., interstitials never separate after encounter. This implies a
reduction of the density of mobile interstitials due to nucleation of
interstitial clusters, i.e., interstitials act as sinks for other
interstitials. However, this creates a population of larger clusters
with a larger crosssection and eventually larger capture numbers. The
reaction kinetics will be affected by the competition of these two
effects.

The rate equations (\ref{rate-irr-eq}) can now be expanded to include
terms representing the above processes, but become even more complex.
In our kMC model, we begin by studying the effect of interstitial
cluster formation by considering completely immobile interstitial
clusters in complete analogy to the vacancy cluster. This situation is
rarely realistic, but provides an upper bound on the magnitude of the
effects. Figure~\ref{intclustimm-fig} shows the evolution of the free
interstitial and vacancy densities as well as the interstitial and
vacancy cluster densities for the same parameters as before. As
expected, the immoblization of interstitials increases the total
number of vacancies in the system. In the final steady state,
$n_{vtot}$ is about three times as large as in the situation in
Fig.~\ref{cluster-fig}. The interstitial clusters (open squares)
nucleate earlier than the vacancy cluster, but their density later
drops below that of the vacancy clusters, since the single
interstitial density is much lower than the single vacancy density.

\begin{figure}[t]
\includegraphics[width=8cm]{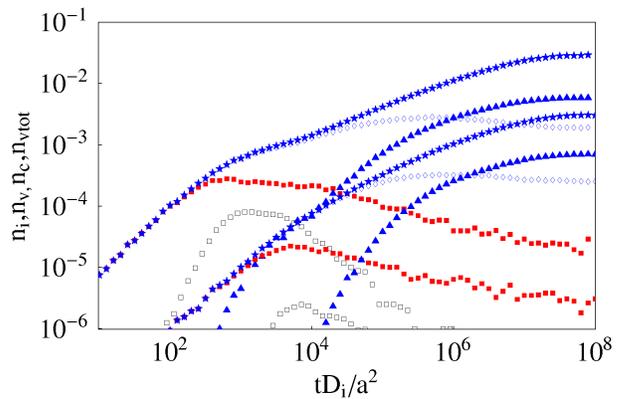}
\caption{\label{intclustmob-fig}Interstitial, vacancy, interstitial
cluster and vacancy cluster densities for $\Gamma=10^5$ and
$\Gamma=10^7$ (symbols as in Fig.~\ref{intclustimm-fig}). Interstitial
clusters $(\square)$ diffuse and rotate with the same rate as single
interstitials. The interstitial cluster density first rises due to
nucleation events, but then rapidly drops as mobile cluster collide
with sinks. }
\end{figure}

The situation in Fig.~\ref{intclustimm-fig} can be favorably
contrasted with that of completely mobile interstitial clusters, i.e.,
the cluster diffuse with the same rates as the single interstitials
regardless of their size. This case is again not fully realistic,
since interstitial clusters seldom rotate into other $\langle 111
\rangle$ directions once they contain several interstitials (i.e.,
they diffusive one-dimensionally \cite{marian01,soneda01}). For small
clusters, however, the completely mobile interstitial cluster case
represents a good approximation. Figure~\ref{intclustmob-fig} shows
the corresponding results for the various desities introduced
above. As in the case of immobile interstitial clusters, the mobile
interstitial cluster density first rises due to nucleation
events. However, the highly mobile interstitial clusters also collide
with the sinks, and the cluster density decreases rapidly. Since the
free interstitial density also declines, nucleation events become
rare. Once steady state has been achieved, the interstitial density
has become so low that the nucleation of new interstitial clusters due
to diffusion is almost completely absent. Consequently, the vacancy
densities and the total swelling rates are the same as for the case of
non-interacting interstitials.

\section{Application to vanadium and iron} 
\label{real-sec}
In this section, we apply our model to the V and $\alpha-$Fe
cases. Vanadium is particularly interesting, because here the effect
of mixed 1D/3D diffusion is most pronounced. First principles
calculations and classical MD simulations of V have yielded estimates
of $\Delta E_i=0.018$eV\cite{zepeda04}, $\Delta E_r=0.44$eV
\cite{zepeda04} and $\Delta E_v\approx 0.5$eV \cite{han1}.  For Fe,
the different ground state of the self-interstitial ($\langle
110\rangle$ instead of $\langle 111\rangle$) leads to a higher
effective activation barrier for 1D migration, $\Delta E_i$, which
ranges between 0.12 eV \cite{marian01} and 0.17 eV
\cite{soneda01}. The rotation barrier was estimated as $\Delta
E_r=0.16$ eV \cite{soneda01}, and the vacancy migration energy is
assumed to be of the same order \cite{soneda03} as in V (the
prefactors for all processes tend to vary by less than an order of
magnitude). One therefore expects that the self-interstitial
trajectories in $\alpha-$Fe will be much more isotropic than in V.

\begin{table}[b]
\begin{center}
\caption{\label{dimless-table}Dimensionless parameters
$\alpha=D_i/D_v$, $\beta=\gamma_r/D_i$,
$\Gamma=(\sqrt{\beta}D_i+D_v)/\sigma F$, and
$\alpha^\prime=\alpha\sqrt{\beta}$ for V and Fe at 300 K and 600 K.}
\begin{tabular}{|c|c|c|c|c|}
\hline
 & V-300K & V-600K & Fe-300K & Fe-600K\\
\hline
$\alpha$ & $10^{8}$ & $10^4$   & $10^{6}$  & $10^{3}$ \\
$\beta$ & $10^{-8}$ & $10^{-4}$ & $10^{0}$  & $10^{0}$  \\
$\Gamma$ & $10^{11}$ & $10^{13}$ & $10^{11}$  & $10^{13}$  \\
$\alpha^\prime$ & $10^4$ & $10^2$ & $10^{6}$ & $10^{3}$  \\ 
\hline
\end{tabular}
\end{center}
\end{table}

The two metals are best compared in terms of the relevant
dimensionless parameter $\alpha$, $\beta$ and $\Gamma$.  Table
\ref{dimless-table} summarizes the values for these quantities using
the above energy barriers and the production rate for ion irradiation
$\sigma F=10^{-3}s^{-1}$ for two representative temperatures $T=300$K
and $T=600$K. We first note that $\Gamma$ is typically larger than
$10^{10}$ and would in fact reach $10^{20}$ when typical production
rates for neutron irradiation are used.  A kMC simulation with such
large values of $\Gamma$ would require very long simulation times,
since diffusion and recombination occur much more often than the
introduction of new defects. It also requires large system sizes,
because the defect densities become very small. In addition, the very
small value of $\beta$ implies very long 1D segments of the
interstitial trajectories between rotations. The period of the kMC
simulation box should be several times larger than the typical length
of those segments in order to properly reproduce the continuum theory
values of the encounter rates. If the box period is much smaller than
the 1D segment, the trajectory will wrap around the box many times,
but is not necessary space-filling.

The last of these issues can be addressed by using a result from
Section \ref{simple-sec}, where we showed that the mixed 1D/3D
encounter rates scale like the ideal 3D rates that are reduced by a
factor $\sqrt{\beta}$. We can therefore replace the explicit 1D/3D
trajectories of the interstitials and interstitial cluster with ideal
3D random walks, but reduce the hopping rate by a factor of
$\sqrt{\beta}$. This procedure leaves the reaction kinetics invariant
(up to small corrections from the capture numbers) and implies that
the effective ratio of time scales between interstitial and vacancy
migration is given by
$\alpha^\prime=D_i\sqrt{\beta}/D_v=\sqrt{D_i\gamma_r/a^2}$. Interestingly,
this ratio is very similar for both V and Fe at 300K and 600K (see
Table \ref{dimless-table}), even though the values of $\alpha$ and
$\beta$ are very different. At a given temperature, the
self-interstitial trajectories in $\alpha$-Fe are much more isotropic
than in V, but the encounter rates with vacancies and sinks only
depend on the product of diffusivity $D_i$ and rotation rate
$\gamma_r$.  Within the present model, one therefore expects that the
point defect reaction kinetics in these two metals is very similar.

\begin{figure}[t]
\includegraphics[width=7.7cm]{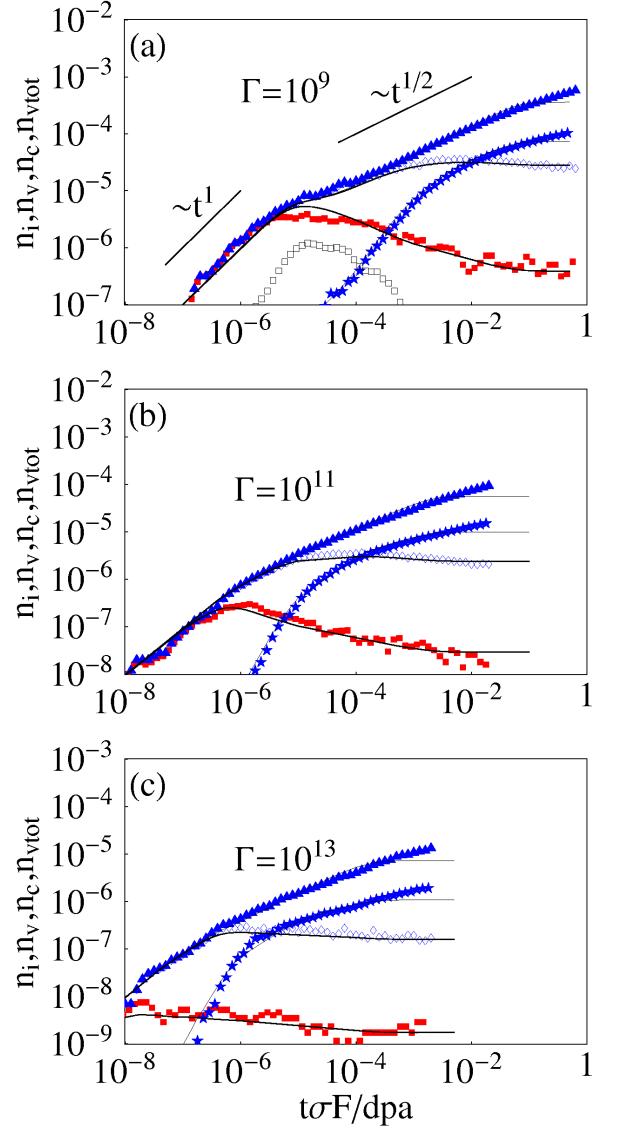}
\caption{\label{van-fig}Interstitial, vacancy, interstitial cluster
and vacancy cluster densities (symbols as in
Fig.~\ref{intclustimm-fig}) with parameters $\alpha^\prime=10^2$,
$n_s=10^{-5}$, for (a) $\Gamma=10^9$ (b) $\Gamma=10^{11}$ and (c)
$\Gamma=10^{13}$ representative for V or $\alpha$-Fe at 600K. Thin
solid lines show the result of numerically integrating
Eqs.~(\ref{rate-irr-eq}) again for $\kappa_m=25$ and $m_{max}=20$. The
solid lines have slope 1 and 1/2, respectively.}
\end{figure}

In Fig.~\ref{van-fig}, we use a ratio of time scales representative
for V/$\alpha$-Fe at $T=600$K and show results for $\Gamma=10^9$,
$\Gamma=10^{11}$ and $\Gamma=10^{13}$.  Here, we have multiplied the
time axis with the production rate, which is a more convential
presentation of the data in experimental studies.  Rescaled by the
production rate, all curves initially coincide and start out with
near-constant slopes. For the present parameters, Regime II is
practically absent, and the vacancy density crosses over immediately
from Regime I (where it rises linearly with dose) into Regime III. The
crossover dose is roughly constant and depends on the sink
density. Note that the growth of $n_{vtot}$ in Regime III is well
described by a $(\sigma Ft)^{1/2}$ power law over several decades
before the final steady-state Regime IV is reached. Since the onset of
Regime IV is independent of time (see Section \ref{simple-sec}),
larger values of $\Gamma$ push the beginning of steady state to
smaller doses. Here, steady state is reached at about 1 dpa for
$\Gamma=10^9$ and earlier for $\Gamma=10^{11}$ and
$\Gamma=10^{13}$. There is some initial nucleation of interstitial
clusters at $\Gamma=10^9$, but as discussed before, all of these
rapidly disappear due to fast collisions with sinks. For higher values
of $\Gamma$, the interstitial cluster density becomes negligibly
small.

In the same figure, we also show the predictions of a numerical
integration of the full set of rate equations (\ref{rate-irr-eq})
using the point cluster model (size independent capture numbers). As
in Fig.~\ref{cluster-fig}, the agreement with the ``exact'' kMC
simulation is satisfactory. Since the computational effort for direct
kMC simulations for $\Gamma>10^{10}$ becomes tremendous, the rate
equation approach, when properly parametrized, is clearly preferable.

\label{van-sec}

\section{Conclusions} 
\label{conc-sec}
We have studied the early-time evolution of point defect populations
with specific reference to migration mechanisms in bcc metals under
constant irradiation conditions. Only very simplified models that
incorporate the most important processes were employed in order to
identify the rate-limiting events. These models were solved using
scaling arguments, direct numerical integration of kinetic rate
equations and full kMC simulations.

In the simplest case, which only considers homogeneous defect
production, recombination and sink absorption, but no interactions
between defects of the same type, the kMC model and the corresponding
rate equations were shown to be in near perfect agreement. We employed
simple random walk arguments to derive a mixed encounter rate that
describes one-dimensional diffusion with occasional rotations. This
encounter rate scales linearly with target density as the isotropic 3D
encounter rate, but is reduced by the square root of the ratio of
rotation rate and hopping rate. As in ref.~\onlinecite{sizmann68},
four distinct scaling regimes for the point defect density with time
where identified. First, the point defect densities increase linearly
in time due to production (Regime I), then saturate as defect
recombination sets in (Regime II). Sink absorption then begins to
reduce the interstitial density and the vacancy density grows in time
with a characteristic $\sim t^{1/2}$ behavior (regime III). A final
steady state (Regime IV) is reached when all loss processes are taken
into account. The full sequence of Regimes I - IV is most visible when
the production rate is not much smaller than the interstitial hopping
rates. Regimes II and III shrink with decreasing production rate, and
in the limit of very small production rates, Regime I crosses over
directly into Regime IV. These results were based on parameters
typical for bcc metals, but the continuum level reaction kinetics
equally applies to fcc metals or other crystal structures provided
they exhibit similar diffusion mechanisms.

The introduction of vacancy reactions to form immobile vacancy
clusters does not change the general sequence of the scaling regimes,
but has a profound effect on the population dynamics. In steady state,
the density of free vacancies and interstitials is reduced relative to
the non-interacting case, but the total number of vacancies $n_{vtot}$
is enhanced. In Regime III, $n_{vtot}\sim t^{1/2}$, and the steady
state Regime IV is reached at later times. Reasonable agreement
between rate equations and kMC could still be achieved in a point
cluster approximation, where the capture numbers are
size-independent. Of particular interest is the size distribution of
vacancy clusters, which grow and shrink under the diffusive arrival of
interstitials and vacancies. For irreversible vacancy aggregation, we
derived a new expression for the cluster size distribution,
$n_c(m)=\kappa_1 n_v^m/\kappa_m(\alpha\sqrt{\beta} n_i)^{m-1}$.  The
general form of this distribution remains when also allowing vacancy
detachment from the cluster.

Immobile interstitial clusters were shown to further increase the
vacancy population. However, interstitial cluster that are as mobile
as single interstitials decay rapidly in relevant parameter
regimes. We therefore expect that nucleation of interstitial clusters
through diffusion plays a negligible role in the microstructural
damage evolution of pure bcc metals. It can become important, however,
if trapping of self-interstitials near impurities occurs.

Because of their importance in applications and the availablitity of
detailed molecular dynamics studies, we applied the model for
parameters suitable for V and $\alpha$-Fe. Within our model, both
metals exhibit similar defect kinetics and therefore similar swelling
rates at a given temperature. For production rates and diffusivities
in typical experimental situations, our calculations revealed a
sequence of growth regimes for the total vacancy density $n_{vtot}$
$\sim t^1$, $\sim t^{1/2}$ and $\sim t^0$. The volume swelling rate
follows the same scaling sequence in the sub-1 dpa
regime. Interestingly, growth of the vacany cluster density as $\sim
t^{1/2}$ over several decades in time has been observed in the much
more detailed kMC simulation of Ref.~\onlinecite{soneda03}, but the
origin of this growth law had not been previously understood. All our
simulations reach the steady state regime at damage levels of less
than 1 dpa. In the experimentally relevant regime of damage levels
between 1 and 100 dpa, it will therefore be more efficient to use
steady state rate equations rather than explicit kMC to predict the
long time point defect distribution evolution.  Nonetheless, it is
important to understand the onset of damage evolution and the
conditions of applicability of the steady state theory.

The fact that experimental swelling rates of V and $\alpha$-Fe are
very different is of course an indication that our present model does
not yet include all relevant phenomena of radiation damage.  One
obvious refinement of the models developed here would be a more
detailed parametrization of the detachment rates (binding energies) of
interstitial and vacancy cluster. However, the intent of the present
study has been to focus on general trends rather than quantitative
comparisons with experiments, and no new physics is expected to appear
from these additional details. For cascade damage conditions
\cite{bullough75}, however, two other processes not included in this
study are known to have a crucial impact on the defect evolution (in
particular void swelling rates).  Frenkel pair production is only
assumed to be homogeneous for energies right above the displacement
threshold, while higher energies lead to the formation of mobile
interstitial and immobile vacancy clusters during the cascade
phase. Our present model is therefore most relevant to low particle
energies. Inclusion of intracascade clustering processes will change
results qualitatively and quantitatively, but requires reliable
information about the cluster size distribution during cascades from
MD simulations.

In addition to this ``production bias'', ``absorption bias'' also
usually exists in bcc metals.  ``Absorption bias'' leads to a
preferred absorption of self-interstitials at sinks and is known to be
an essential driving force for swelling. The origin of this bias,
which leads to increased capture numbers, are long range elastic
interactions between point defects and sinks such as dislocations
\cite{kamiyama94,bullough70} or grain boundaries \cite{samaras03}.
Extentions of the model to include these effects should provide a
fruitful topic for future work. The combination of kMC and rate theory
can then be used to determine which parameters are most important in
radiation damage, so that atomistic simulation resources can be better
focused.

The authors gratefully acknowledge useful discussions with
L.~Zepeda-Ruiz, B.~Wirth, and N.~Ghoniem as well as the support of the
Office of Fusion Energy Sciences (Grant DE-FG02-01ER54628).


\end{document}